\documentclass{%
JHEP3}

\usepackage{graphicx}

\def\0{\over } \def\2{{1\over2}} \def\4{{1\over4}}
\def\5{\hat } \def\6{\partial }

\def\({\left(} \def\){\right)} \def\<{\langle } \def\>{\rangle }

\newcommand{\bea}{\begin{eqnarray}}
\newcommand{\eea}{\end{eqnarray}}
\newcommand{\be}{\begin{equation}}
\newcommand{\ee}{\end{equation}}
\newcommand{\nn}{\nonumber\\ }
\newcommand \beq{\begin{eqnarray}}
\newcommand \eeq{\end{eqnarray}}

\def\Im{{\,\rm Im\,}}
\def\Re{{\,\rm Re\,}}
\def\tr{{\,\rm tr\,}}

\preprint{ECT-06-22, SPhT-T06/163, TUW-06/07 
}

\title{Hard thermal loops and the entropy of supersymmetric Yang-Mills theories}
\author{J.-P. Blaizot\\
ECT*, Villa Tambosi, Strada delle Tabarelle 286,\\
I-38050 Villazzano Trento, Italy}
\author{E. Iancu\\
Service de Physique Th\'eorique, CE Saclay,
        F-91191 Gif-sur-Yvette, France}
\author{U. Kraemmer and A. Rebhan\\
Institut f\"ur Theoretische Physik, Technische
Universit\"at Wien, \\Wiedner Hauptstr.~8-10, A-1040 Vienna,
Austria }

\abstract{%
We apply the previously proposed scheme of approximately
self-consistent hard-thermal-loop resummations in the entropy of
high-temperature QCD to ${\cal N}=4$ supersymmetric Yang-Mills
(SYM) theories and compare with a (uniquely determined)
$R_{[4,4]}$ Pad\'e approximant that interpolates accurately
between the known perturbative result and the next-to-leading
order strong-coupling result obtained from AdS/CFT correspondence.
We find good agreement up to couplings where the entropy has
dropped to about 85\% of the Stefan-Boltzmann value.
This is precisely the regime which in
purely gluonic QCD corresponds to temperatures above 2.5
times the deconfinement temperature and for which this method of 
hard-thermal-loop resummation
has given similar good agreement with lattice QCD results. 
This suggests that in this regime the entropy of both QCD 
and $\mathcal N=4$ SYM
is dominated by effectively weakly coupled 
hard-thermal-loop quasiparticle degrees of freedom. In $\mathcal N=4$ SYM,
strong-coupling contributions
to the thermodynamic potential take over when the entropy drops below
85\% of the Stefan-Boltzmann value.
}


\begin{document}

\section{Introduction}

Much theoretical effort is presently devoted to the study of QCD in a
regime where it appears strongly coupled. This effort is mainly
triggered by the beautiful data obtained at the Relativistic Heavy Ion
Collider, which reveal that matter produced in high energy
nucleus-nucleus collisions is strongly interacting, and may, to a good
approximation, be described as an ideal fluid with very low ratio of
shear viscosity over entropy density $\eta/\mathcal S$ (see, e.g.,
\cite{Gyulassy:2004zy+Shuryak:2004cy+Muller:2006ee} and references therein).

At the same time, new techniques have become available that allow the
treatment of strongly coupled gauge theories.  In particular,
$\mathcal N=4$ supersymmetric Yang-Mills theories (SYM) in the limit
of large number of colors $N$ and strong 't Hooft coupling
$\lambda\equiv g^2 N$ is now often taken as a model for hot QCD.
Through the AdS/CFT conjecture \cite{Maldacena:1997re,Aharony:1999ti} one
can make predictions for the otherwise inaccessible strong-coupling
regime, in particular real-time quantities (such as transport
coefficients) \cite{Herzog:2006gh,Casalderrey-Solana:2006rq,Liu:2006ug,Janik:2006gp,Caron-Huot:2006te},
where lattice is fraught with large uncertainties
\cite{Aarts:2002cc}. While $\eta/\mathcal S \sim
(\lambda^2\log\lambda)^{-1}\gg 1$ at weak coupling
\cite{Arnold:2000dr+Arnold:2003zc,Huot:2006ys}, the AdS/CFT correspondence
gives $\eta/\mathcal S=1/4\pi+O(\lambda^{-3/2})$ at large 't Hooft
coupling \cite{Policastro:2001yc,Buchel:2004di}, which has been
argued to be a universal result in strongly coupled gauge theories
with gravity duals \cite{Buchel:2003tz,Kovtun:2004de,Benincasa:2006fu}.

In \cite{
Gubser:1998nz} the thermodynamic entropy of $\mathcal N=4$ SYM
theory at large 't Hooft coupling and large $N$ has been
determined from the AdS/CFT correspondence as%
\footnote{In \cite{Gubser:1998nz} as well as
Refs.~\cite{Fotopoulos:1998es,Landsteiner:1999gb,Nieto:1999kc,Buchel:2004di}
this result appears with $(2\lambda)^{-3/2}$
in place of $\lambda^{-3/2}$, which corresponds to a definition
of the coupling that differs from the (universally adopted)
standard usage of $g$ at weak coupling. The latter requires
$\lambda\equiv g^2N=L^4/\alpha'{^2}=4\pi g_{\rm string}N$, where $L$ is the
curvature scale of $AdS_5$, $\alpha'$ the string tension, and
$g_{\rm string}$ the string coupling constant
 (see also \cite{Gubser:2006qh}).
The comparison of weak and strong coupling results in 
Refs.~\cite{Fotopoulos:1998es,Landsteiner:1999gb,Nieto:1999kc,Buchel:2004di}
(with the exception of \cite{Aharony:1999ti})
is therefore incorrect; Ref.~\cite{Kim:1999sg} does have
compatible weak and strong coupling results (apart from
mixing $\lambda$ with $\lambda^2$), but quotes a then incompatible
relation between $\lambda$ and $g_{\rm string}$.
}
\be 
\mathcal S/\mathcal S_0=
{3\04} \( 1 + {15\zeta(3)\08}\lambda^{-3/2} +\ldots\)
\label{scsym}. 
\ee
The fact that, in lattice QCD, the value of
$\mathcal S/\mathcal S_0$ for temperatures above $3T_c$ appears to
be comparable to this strong-coupling result for $\mathcal N=4$ SYM theory
has led to the further suggestion
that also for such temperatures QCD is still in a strong-coupling
regime. While in this range of temperature the violation of scale
invariance in QCD is negligible (in the sense that
$(\epsilon-3P)/\epsilon$ is fairly small), and therefore does not
spoil the direct comparison with (conformally invariant) $\mathcal
N=4$ SYM theory, we shall argue that such a comparison rather
suggests a description in terms of effectively weakly coupled
degrees of freedom for both theories.

That such a description can be successful should not be taken as
implying that the system as a whole is literally weakly coupled
(typically $\lambda\gtrsim 1$), nor that its properties can be
obtained through a straightforward expansion in powers of
$\lambda$. In fact, the weak-coupling expansion for the
thermodynamical potential of $\mathcal N=4$ SYM theory is
presently known to order $\lambda^{3/2}$
\cite{Fotopoulos:1998es,Vazquez-Mozo:1999ic,Kim:1999sg,Nieto:1999kc},
\be
 \mathcal S/\mathcal S_0=1-{3\02\pi^2}\lambda+{\sqrt2+3\0\pi^3}\lambda^{3/2}
+\ldots
\ee
but the extremely poor convergence of this expansions seems to
limit its usefulness to very small values of the coupling, where
the deviation from the ideal-gas result is only of the order of a few
percent (see Fig.~\ref{figsym} below). 
In particular, this precludes any comparison with the
corresponding results at strong coupling. This situation is
indeed very much like in real QCD, where the perturbative series
for the thermodynamic potentials of hot QCD \cite{%
Arnold:1995eb,Zhai:1995ac,Braaten:1996jr,Kajantie:2002wa} shows
convergence only for temperatures more than $10^5$ times the
deconfinement temperature where the coupling is so small that
deviations from ideal-gas thermodynamics are minute.

However, the same impasse appears also in such simple
theories as unbroken massless O($N$) $\phi^4$ theories, where in
the limit $N\to\infty$ the entropy is given by an interaction-free
expression for quasi-particles with thermal masses determined by a
one-loop gap equation \cite{Drummond:1997cw}, while the
corresponding perturbative expansion is ill behaved except for
very small values of the (scalar) 't Hooft coupling. As this
example shows, the failure of finite-temperature perturbation
theory in nonabelian gauge theories is not necessarily due to
specifically nonabelian nonperturbative effects, but largely due
to an incomplete resummation of screening effects (at least as far
as thermodynamical quantities dominated by hard degrees of
freedom, such as the entropy, are concerned).\footnote{Of course,
nonabelian gauge theories do have specific nonperturbative phenomena
such as confinement in the chromomagnetostatic sector, characterized
by the scale $g^2 T$. Depending on the quantity under consideration,
the latter can show up in rather low orders of perturbation 
theory \cite{Rebhan:1993az},
but in the thermodynamic quantities pressure or entropy, they
appear only at 4-loop order, parametrically suppressed by 6 powers of
the coupling $g$. We are primarily interested in locating the regime where
a weak-coupling expansion makes sense (sufficiently above
the deconfinement transition in the case of QCD) and where we expect
such contributions of order $g^6$ to be comparatively small.}  For scalar
theories, a corresponding reorganization of perturbation theory
has been proposed in Refs.~\cite{Karsch:1997gj,Andersen:2000yj}
using simple effective
mass terms, and generalized to gauge theories in Refs.~\cite{%
Andersen:1999fw+Andersen:1999sf+Andersen:1999va,%
Andersen:2002ey} by replacing the simple mass term by the
gauge-invariant hard-thermal-loop (HTL) effective action
\cite{Braaten:1992gm,Frenkel:1992ts,Blaizot:2001nr}.  However, this
method generally suffers from artificial ultraviolet divergences and
moreover does not take into account that the HTL effective action is
applicable at hard momentum scales only for small virtuality.

In Refs.~\cite{Blaizot:1999ip,Blaizot:1999ap,Blaizot:2000fc}, three of us have
proposed a different scheme to resum HTL effects which is manifestly
ultraviolet finite and applies the (non-local) HTL corrections only in
kinematical regimes where these are applicable at weak coupling.  When
applied to QCD, we have found reasonable agreement with available
lattice data (see Fig.~\ref{figqcd}) for temperatures down to about 3
times the deconfinement temperature\footnote{Similar agreement can be
  obtained by suitable renormalization-scale optimizations of the
  three-loop result from dimensional reduction
  \cite{Blaizot:2003iq,Ipp:2003yz}.}, suggesting that there the entropy of the
quark-gluon plasma can be understood in terms of weakly interacting
HTL quasiparticles, the main effect of the interactions being
incorporated into a renormalization of the quasiparticle properties.

In the following, we shall recapitulate our HTL resummation scheme
for the entropy and extend our results to SYM theories, where they
can be compared with the strong-coupling result of
Eq.~(\ref{scsym}). In Ref.~\cite{Blaizot:2005wr}, we have recently
tested our approach successfully in the exactly solvable case of
large-$N_f$ gauge theories, which is however essentially Abelian.
The results from the AdS/CFT correspondence deal with the other
extreme of infinite number of colors.

\section{Quasiparticle entropy and HTL resummation}

Expressed in terms of (self-consistently)  dressed propagators
($D$ and $S$ for bosonic and fermionic ones) and dressed
self-energies ($\Pi$ and $\Sigma$, respectively), the entropy
admits the following (exact) representation
\cite{Vanderheyden:1998ph,Blaizot:1999ap,Blaizot:2000fc}:
 \bea
\label{S2loop} &&{\mathcal S}=-\tr
\int{d^4k\0(2\pi)^4}{\6n(\omega)\0\6T} \Bigl[ \Im \log D^{-1}-\Im
\Pi \Re D \Bigr] \nn &&-2\tr
\int{d^4k\0(2\pi)^4}{\6f(\omega)\0\6T} \Bigl[ \Im \log S^{-1}-\Im
 \Sigma \Re S \Bigr] + \mathcal S'
\eea
($n$ and $f$ are Bose-Einstein and Fermi-Dirac distributions,
respectively), where $\mathcal S'$ is a 3-loop order quantity
that, loosely speaking, describes residual interactions of the
quasiparticles:  the bulk of the interactions  have been
incorporated in the spectral data determining the  properties of
these quasiparticles through the dressed propagators.

In practice, some approximations are needed in order to
evaluate the quantities which appear in Eq.~(\ref{S2loop}). Our approximation scheme is based on a skeleton expansion truncated at two loop order, whereby ${\cal S}'=0$. Furthermore, the exact expressions for the self-energies that would result from solving self-consistent gap equations are approximated by their hard-thermal loop expressions, in a way that is described below. This further  approximation has the virtue of leading to gauge invariant results, which would otherwise not be the case in such an approach.

In perturbation theory, HTL self energies provide the relevant
leading-order corrections for soft momenta $\omega,k\sim g T$ as
well as for hard momenta, provided the latter have small
virtuality. One of the merits of the entropy expression
(\ref{S2loop}) in comparison with the usual loop-wise expansion of
the pressure (whose first derivative with respect to temperature
also yields the entropy) is that the contributions of order
$g^2$ and $g^3$ are generated by either soft momenta or hard
momenta with small virtuality, for which the HTL approximation is
meaningful.

Indeed, to leading order in $g^2$, all interaction effects
are encoded in the (asymptotic) thermal masses of hard excitations
and summarized by the simple formula
\cite{Blaizot:1999ap,Blaizot:2000fc} 
\be\label{S2} \mathcal
S_2=-T\left\{\sum_b {m_{\infty (b)}^2\012} + \sum_f {m_{\infty
(f)}^2\024} \right\}, 
\ee 
where the sum is over all bosonic and
fermionic degrees of freedom. The asymptotic thermal masses can
be read from the
HTL self energies, for which they provide the scale and which
read explicitly
\bea
\label{HTLPiT}
\hat\Pi_T&=&m_{\infty (g)}^2  + {\omega^2-k^2\02k^2}\Pi_L,\\
\label{HTLPiL}
\hat\Pi_L&=&2 m_{\infty (g)}^2 \left(1-{\omega\02k}\log {\omega+k\0\omega-k}
\right)
\eea
for the spatially transverse and longitudinal gauge bosons and
\be\label{HTLSigma}
\hat\Sigma_{\pm}={m_{\infty (f)}^2\02k}\left(
1-{\omega\mp k\02k}\log {\omega+k\0\omega-k}
\right)
\ee
for the parts of the fermionic self energy with chirality equal/opposite to
helicity,
while for scalars there is no momentum dependence, $\hat\Pi_s\equiv
m_{\infty (s)}^2$.

Specializing to unbroken SYM theories (without matter in the
fundamental representation), for the possible choices of $\mathcal
N$ we have a number of adjoint scalar ($n_s$) and fermionic
($n_f$) degrees of freedom in addition to the two polarizations of
the gauge bosons as given by Table I ($\mathcal N$=0 corresponding
to non-supersymmetric pure-glue QCD). The asymptotic thermal
masses of the gauge bosons, adjoint scalars, and adjoint fermions
turn out to be given by
\bea
m_{\infty (s)}^2=
m_{\infty (g)}^2&=&{2+n_s+n_f/2\012}\lambda T^2\label{minftyb},\\
m_{\infty (f)}^2&=&{2+n_s\08}\lambda T^2, \label{minftyf}
\eea
which coincide when
supersymmetry is realized (whereas only $m_{\infty (g)}$ is
present anyway when $\mathcal N$=0).

Writing
\be\label{Spt}
\mathcal S/\mathcal S_0=1+a_2 {\lambda\0\pi^2}+
a_3 {\lambda^{3/2}\0\pi^3}+\ldots,
\ee
the coefficient $a_2$ is determined by these results for the
asymptotic thermal masses through (\ref{S2}) as tabulated in
Table \ref{tablem} for the possible values of $\mathcal N$.

\TABLE{
\caption{Number of adjoint scalar ($n_s$) and fermionic ($n_f$)
degrees of freedom for different $\mathcal N$, and the resulting
common asymptotic thermal mass for all excitations. Also given are
the effective numbers of freedom as measured by $g_*\equiv
\mathcal S_0/(2\pi^2T^3/45)$; the coefficients $a_2$ and $a_3$
in the perturbative expression for $\mathcal S/
\mathcal S_0$; and the values $\lambda=\lambda_*$ where $\mathcal
S_3=|\mathcal S_2|$. \label{tablem}}
\begin{tabular}{c||c|c|c|c|c|c|c}
$\mathcal N$ & $n_s$ & $n_f$ & $m_\infty^2/\lambda T^2$ & $g_*/N_g$ & $a_2$ & $a_3$ & $\lambda_*$ \\
\hline\hline
0 & 0 & 0 & 1/6 & 2 & $-5/16$ & $5/4\sqrt{3}$ & 1.85\\
1 & 0 & 2 & 1/4 & 15/4 & $-3/8$ & $1/\sqrt{2}$ & 2.78\\
2 & 2 & 4 & 1/2 & 15/2 & $-3/4$ & $1+1/\sqrt{2}$ & 1.91\\
4 & 6 & 8 & 1 & 15 & $-3/2$ & $3+\sqrt{2}$ & 1.14\\
\end{tabular}
}

At low momenta $p\lesssim \lambda^{1/2}T$, the degeneracy of the
thermal masses of the various excitations is broken, but to
leading order they are determined by the structure of the HTL
effective action in a universal fashion. This gives different
momentum dependent HTL masses for gauge bosons and fermions, while
the HTL mass of scalar particles is momentum independent. For
gauge bosons, the static Debye mass differs from the asymptotic
thermal mass according to $m_D^2=2 m^2_{\infty(g)}$, while the
plasma frequency in the gauge boson propagator is given by
$\omega_{pl}^2={2\03}m^2_{\infty(g)}$ and the fermionic plasma
frequency by $M_f^2={1\02}m^2_{\infty(f)}$.

Perturbation theory at finite temperature is made ill-behaved by the
appearance of so-called plasmon terms $\propto \lambda^{3/2}$,
reflecting the existence of collective phenomena 
\be {\mathcal
  S_3\0N_g}={1\03\pi}\left\{m_D^3+n_s m_{\infty (s)}^3\right\}
={2\sqrt2+n_s\03\pi}m_\infty^3, \ee 
where $N_g=N^2-1$. 
For the different $\mathcal N$ this yields the
coefficients $a_3$ in (\ref{Spt}) as tabulated in Table \ref{tablem}.
The plasmon term renders $\mathcal S/\mathcal S_0$ a nonmonotonic
function of $\lambda$, which after going through a minimum
attains again the Stefan-Boltzmann value at the
uncomfortably low values $\lambda=\lambda_*$ 
as given in Table \ref{tablem}, 
the minimum occurring at $\lambda={4\09}\lambda_*$. Clearly,
strict perturbation theory breaks down for such couplings.


However, as the example of O($N\to\infty$) $\lambda\phi^4$ theory
shows \cite{Blaizot:2003tw}, such a breakdown of perturbation theory
even occurs in an essentially  free theory for quasiparticles
($\mathcal S'=0$
in this theory) when the coupling constant dependence of the entropy
expression (\ref{S2loop}) through thermal masses is expanded out. 
In Refs.~\cite{Blaizot:1999ip,Blaizot:1999ap,Blaizot:2000fc}
we have therefore proposed to
evaluate the thermal self energies in HTL theory to leading and
next-to-leading order and to keep the entropy
expression in Eq.~(\ref{S2loop}) unexpanded in
the coupling constant.

Approximating thus $\Pi$ and $\Sigma$ in (\ref{S2loop}) by the
leading-order HTL expressions (\ref{HTLPiT})--(\ref{HTLSigma})
turns out to take into account all of
$\mathcal S_2$. However, only a quarter of the plasmon term $\mathcal
S_3$ is due to the entropy of the soft excitations in the HTL
approximation, the remaining three quarters arise from next-to-leading
order corrections of the asymptotic masses of the various excitations.
These corrections, which are due to the coupling of the hard
  degrees of freedom with the soft collective excitations,  are
momentum dependent even for hard momenta $k\sim T$, but to order
$\lambda^{3/2}$ only their average, weighted according to \be\label{deltamas}
\bar\delta m_{\infty}^2={\int dk\,k\,n'(k) \Re \delta\Pi(\omega=k) \0
  \int dk\,k\,n'(k)} ,\ee is relevant (with a similar expression for
the fermionic case).

For the minimally supersymmetric case ($\mathcal
N=1$), the solution to the asymptotic thermal masses can be taken over
from Ref.~\cite{Blaizot:1999ap}, Eq.~(13), by replacing Casimir
factors of the fundamental representation by those of the adjoint one.
This shows that the degeneracy of the asymptotic thermal masses in
supersymmetric theories carries over to $m^2_\infty+\bar \delta
m_\infty^2$. Generalizing to $\mathcal N>1$ and $n_s\not=0$, we find
\be\label{bardeltaminfty} \bar \delta m_\infty^2=-\lambda T m_\infty
{2\sqrt2+n_s\04\pi}, \ee
which is valid for all the cases covered by
Table I.

Interpreted strictly perturbatively, this leads to tachyonic
masses for $\lambda \gtrsim \lambda_*$. However, in scalar
O($N\to\infty$) models, where the same difficulty occurs,
approximating the true gap equation by one that is quadratic in $m$
gives surprisingly accurate results even for very large coupling.
This just corresponds to putting
\be\label{bardeltaminfty_GAP}
m_\infty^2= m^2_\infty|_{\rm HTL}
            -\lambda T m_\infty {2\sqrt2+n_s\04\pi}, \ee
where the notation $m^2_\infty|_{\rm HTL}$ is now used for the
leading-order quantities in Eqs.\ (\ref{minftyb})--(\ref{minftyf}),
and $m_\infty$ is defined as the solution to the above equation.%
\footnote{Since this procedure may look rather {\it ad-hoc} at this
point, let us emphasize that, in the case of the scalar theory at
least, it corresponds to a meaningful approximation to the actual
gap equation required by the condition of self-consistency
\cite{Blaizot:2000fc}.}

In Ref.~\cite{Blaizot:2000fc} we have 
proposed to employ such quadratic gap equations for correcting the mass
of hard excitations
by using (\ref{bardeltaminfty_GAP}) as the prefactor in
the HTL self energies (\ref{HTLPiT}) and (\ref{HTLPiL}), but only
when the momentum is hard as specified by a dividing scale
$\Lambda=\sqrt{2\pi T m_D c_\Lambda}$. This is to take into account that
soft excitations are
known to have rather different corrections in gauge theories (e.g.\,
the Debye mass squared at next-to-leading order receives positive and
moreover logarithmically enhanced corrections to order $\lambda^{3/2}$
\cite{Rebhan:1993az}, which equally applies for the adjoint scalars%
\footnote{The term proportional to $\log\lambda$ in the NLO correction
  to the adjoint scalars' screening mass is in fact identical to that
  of the Debye mass, which in the static case is likewise carried by an
  adjoint scalar, namely $A_0$.}). The introduction of $\Lambda$ means
that for soft excitations the leading-order HTL masses are kept
untouched%
\footnote{Repeating the procedure of
  Ref.~\cite{Blaizot:2000fc}, the separation scale is only introduced
  for bosonic modes which are solely responsible for the plasmon
  effect.}. As part of the theoretical error estimate, $c_\Lambda$ is
being varied in the range $\2\ldots 2$. In QCD, an even greater
uncertainty comes however from the dependence of the coupling on the
renormalization scale $\bar\mu_{\rm MS}$ whose central value we choose
as $2\pi T$, which is close to the optimal value considered in
Ref.~\cite{Laine:2005ai}.

\section{Results and discussion}

\FIGURE{
\includegraphics[width=0.63\textwidth]{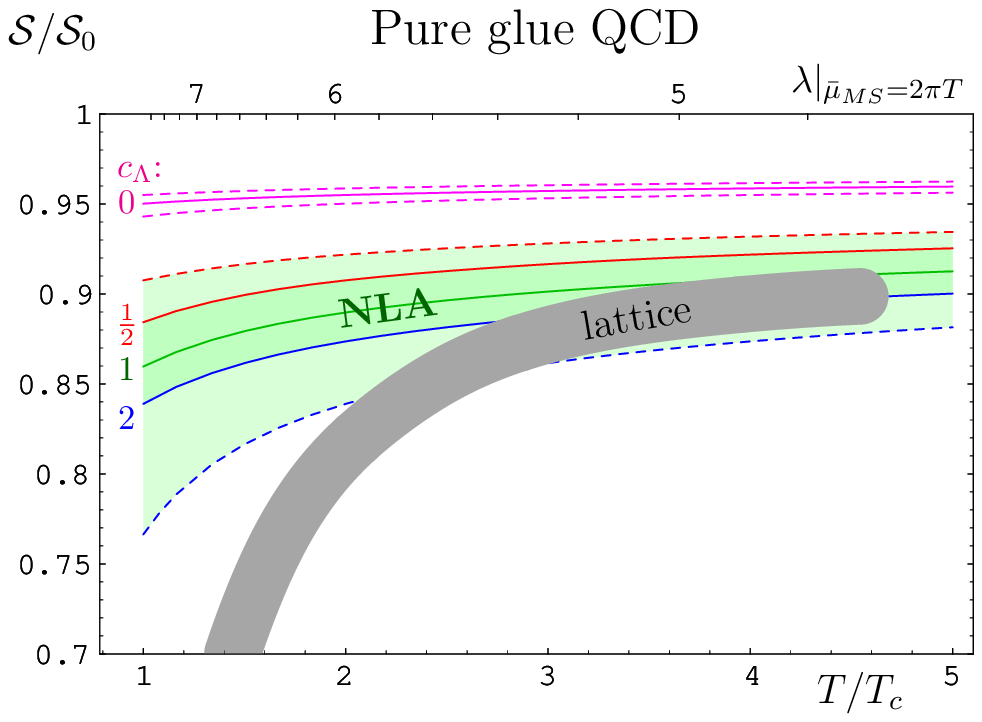}
\caption{The HTL resummed entropy density of pure glue QCD,
divided by its Stefan-Boltzmann value as calculated in a
next-to-leading approximation (NLA) according to
Ref.~\cite{Blaizot:2000fc}. The full lines are evaluated with
renormalization scale $\bar\mu_{\rm MS}=2\pi T$ and different
$c_\Lambda$ (see text).
The upper and lower
dashed lines correspond to renormalization scales $\bar\mu_{\rm
MS}=4\pi T$ and $\pi T$, respectively. The dark band represents
the continuum-extrapolated lattice results of
Ref.~\cite{Boyd:1996bx}. \label{figqcd}}
}

For comparison with our new results for $\mathcal N=4$ SYM which
we display further below, we start by exhibiting the results of
our corresponding previous calculation in pure glue QCD. In
Fig.~\ref{figqcd} the result obtained in
Ref.~\cite{Blaizot:2000fc} is reproduced, where the band marked
``NLA'' corresponds to the HTL entropy evaluated to
next-to-leading order according to the above procedure with
$c_\Lambda=\2\ldots 2$ and $\bar\mu_{\rm MS}=2\pi T$. In the
larger band bounded by dashed lines, the renormalization scale is
also varied, $\bar\mu_{\rm MS}=\pi T\ldots 4\pi T$. (The lines
marked $c_\Lambda=0$ correspond to also rescaling the HTL masses
of soft excitations by the same factor as the asymptotic thermal
mass, which, as we argued above, does not do justice to the
former.) These results are compared with the
continuum-extrapolated lattice results of Ref.~\cite{Boyd:1996bx}
where the width of the dark band is meant as a rough indicator of
the lattice errors (the more recent results of
Ref.~\cite{Okamoto:1999hi} would be centered about the upper limit
of this band). For temperatures larger than about 3 times the
phase transition temperature $T_c$ one finds a remarkably
good agreement between our NLA approximation and the lattice
results, which is perhaps even better than one would have the
right to expect (given the relatively large values of the
corresponding coupling `constant', $g^2_{\rm QCD}=\lambda/3$, with
$\lambda|_{\bar\mu_{\rm MS}=2\pi T}$ as indicated on the top of
Fig.~\ref{figqcd}).

The strong-coupling results provided by AdS/CFT correspondence now
give us an opportunity to test our procedure in the case of
$\mathcal N=4$ SYM theory.
While no lattice results are available, the weak-coupling and
the strong-coupling results to order $\lambda^{3/2}$ and $\lambda^{-3/2}$,
respectively, together should give an idea of the true result.
To make this a bit more quantitative, we shall construct a Pad\'e approximant%
\footnote{Pad\'e approximations have already been considered in
Ref.~\cite{Kim:1999sg}, however only for the weak-coupling part.
The two variants considered in Ref.~\cite{Kim:1999sg} turn out to
coincide numerically rather closely with the NLA results
for $c_\Lambda=1$ and $1/2$ (see Fig.~\ref{figsym} for these latter results).}
which reproduces both, the weak-coupling and
the strong-coupling results, to the order they are known
by a rational function of $\lambda^{1/2}$ or, equivalently $\lambda^{-1/2}$.
The fact that at both zero and infinite coupling a finite result is obtained
requires that this rational function has polynomials of equal degree in
numerator and denominator. It turns out that there is just sufficient
information to fix uniquely all the coefficients in the $R_{[4,4]}$ approximant
\be\label{R44}
R_{[4,4]}=
{1+\alpha \lambda^{1/2}+\beta\lambda+\gamma\lambda^{3/2}+\delta\lambda^2 \0
1+\bar\alpha \lambda^{1/2}+\bar\beta\lambda+\bar\gamma\lambda^{3/2}+\bar\delta\lambda^2 }.
\ee

\FIGURE[t]{
\includegraphics[width=0.62\textwidth]{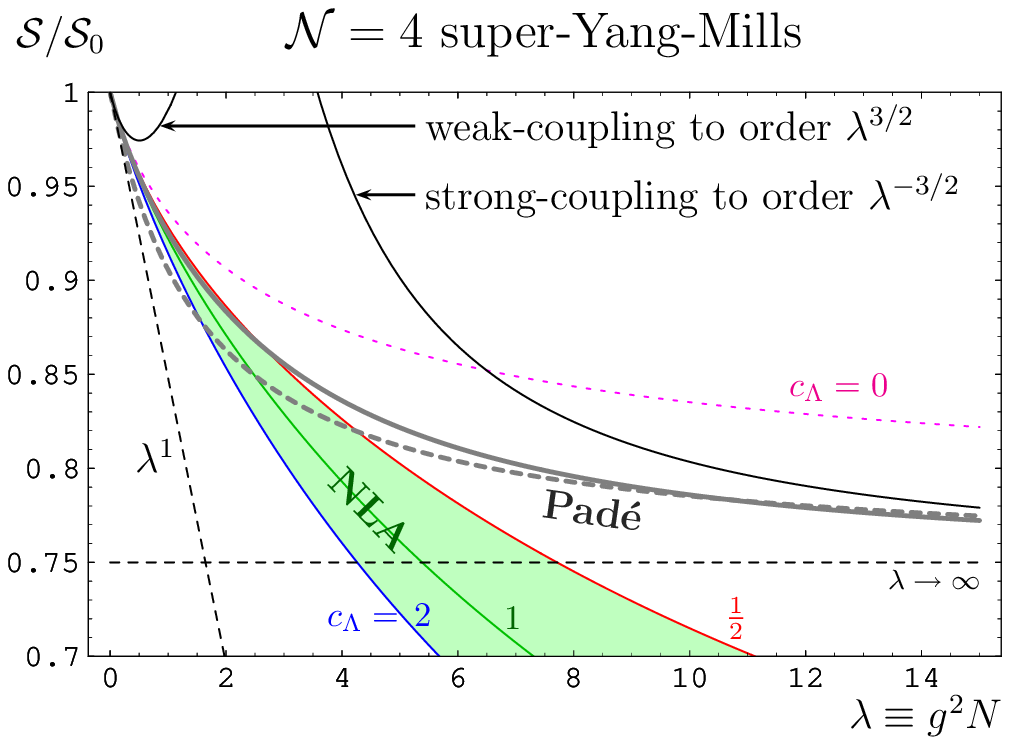}
\caption{Weak and strong coupling results for the entropy density
of $\mathcal N=4$ SYM theory together with the NLA results
obtained in analogy to QCD (cp.\ Fig.\ \ref{figqcd}), but as a
function of $\lambda$, which here is a free parameter.
The dashed and full heavy gray lines represent the Pad\'e
approximants $R_{[1,1]}$ and $R_{[4,4]}$ which interpolate
between weak and strong coupling results to leading and next-to-leading
orders, respectively.
\label{figsym}}
}

The fact that the weak coupling result has no term of order
$\lambda^{1/2}$, while the strong-coupling result approaches $3/4$
at large couplings, where it has no terms of order
$\lambda^{-1/2}$ or $\lambda^{-1}$, completely constrains the
coefficients in the denominator of (\ref{R44}) in terms of those
in the numerator:
 \be
 \bar\alpha=\alpha,\quad \bar\beta={4\03}\beta,\quad
 \bar\gamma={4\03}\gamma,\quad \bar\delta={4\03}\delta. \ee
The remaining four independent coefficients are then uniquely
fixed by the results quoted above for the weak-coupling expansion to
order $\lambda^{3/2}$ and, respectively, the strong-coupling
expansion to order $\lambda^{-3/2}$, together with the expectation
that the subsequent term in the latter should be suppressed by at
least a factor of $\lambda^{-1}$. One thus finds
 \bea
\alpha&=&{2(9+3\sqrt2+\gamma\pi^3)\09\pi},\quad
\beta={9\02\pi^2},\quad\nn \gamma&=&{2\015\zeta(3)},\quad
\delta={2\015\zeta(3)}\alpha\,.
\eea Gratifyingly, the resulting rational function has no pole at positive
values of $\lambda$ and it interpolates monotonically between zero
and infinite coupling, as 
commonly expected \cite{Aharony:1999ti}.
In Fig.~\ref{figsym},
the Pad\'e approximant (\ref{R44}) to the weak- and
strong-coupling results is represented by the heavy gray line
marked ``Pad\'e''. In order to get an idea of the convergence of
successive Pad\'e approximants, the simpler Pad\'e approximant
$R_{[1,1]}$ which only reproduces the weak-coupling result to
order $\lambda^1$ and the constant limit at infinite
coupling is also shown (dashed gray line). This comparison
suggests that the true result should be not too far from the
$R_{[4,4]}$ result, which we thus adopt for assessing the
HTL-resummed (NLA) entropy discussed above. Evaluating the latter
precisely in the same way as we did in QCD (but with less
uncertainty, since because of the finiteness of the $\mathcal N=4$ 
SYM theory there is no ambiguity from renormalization-scale dependence),
we obtain a result that for\footnote{%
Just as in the case of QCD we find that it is indeed crucial that
the correction (\ref{bardeltaminfty}) is applied only for hard excitations,
as we have done by introducing a separation of
hard and soft physics by the scale $\Lambda=\sqrt{2\pi T m_D c_\Lambda}$
and $c_\Lambda\sim 1$, which we take as 
validation of our procedure proposed in the case of QCD. Note that for
this separation to make sense it is
essential that $m_D\ll 2\pi T$, i.e.,
$\lambda \ll 2\pi^2\sim 20$, which is indeed fulfilled in
the range where our NLA result is plotted for $\mathcal N=4$ SYM.}
$c_\Lambda$ between $1/2$ and 2 agrees remarkably well with the
Pad\'e results up to a coupling $\lambda\sim 3$, whereas the ordinary
weak-coupling result to order $\lambda^{3/2}$ fails for all $\lambda\gtrsim 0.5$.

After this comparison within $\mathcal N=4$ SYM, we may also try to
compare the situation between the latter theory and QCD.
At the top of Fig.~\ref{figqcd} we have indicated the values of $\lambda=3g^2$
for (purely gluonic)
QCD, when renormalized with two-loop running coupling at the scale
$\bar\mu_{MS}=2\pi T$ and for $T_c=1.14 \Lambda_{\rm QCD}$.
In Refs.~\cite{Chesler:2006gr,Caron-Huot:2006te,Huot:2006ys} (see also
Ref.~\cite{Gubser:2006qh}), where recently other weak and strong
coupling results of $\mathcal N=4$ SYM theory have been compared, it
has been argued that
values of $\lambda$ in QCD ($\lambda_{\rm QCD}$) and in ${\cal
N}=4$ SYM ($\lambda_{\rm SYM}$) should be related such that
comparable thermal effects like screening masses arise. Now,
(leading-order perturbative) thermal masses in $\mathcal N=4$ SYM
theory are 6 times as large as in purely gluonic QCD, just
reflecting the much higher number of degrees of freedom for a
given color number $N$. So, by itself, this comparison would
suggest a factor of 6 of difference between $\lambda_{\rm QCD}$
and $\lambda_{\rm SYM}$.
However, instead of the asymptotic thermal masses, one
could also compare the normalization of
$\mathcal S_0$, the ratio $\mathcal S_2/\mathcal S_0$ or $\mathcal
S_3/\mathcal S_2$ as done in Table I, so that the reduction of
$\lambda_{\rm SYM}$ over $\lambda_{\rm QCD}$ could be argued to be
anywhere between  $1/6$ and $1/1.6$.

In fact, the lattice results for the entropy of (pure glue)
QCD and the Pad\'e estimate of the entropy of SYM present yet
another possibility for such a comparison, and to that aim we
shall focus our attention at larger temperatures in QCD, where
$\mathcal S/\mathcal S_0$ flattens out. (Close to the phase
transition the entropy of pure glue QCD becomes very small,
whereas $\mathcal S/\mathcal S_0$ in $\mathcal N=4$
SYM approaches 3/4 at very large coupling.) In QCD,
$\mathcal S/\mathcal S_0$ flattens out for $T\ge 3T_c$,
corresponding to $\lambda_{\rm QCD}\le 5.3$. The pure glue 4D
lattice results, which are available up to about $5T_c$, have
$\mathcal S/\mathcal S_0$ interpolating between about $0.87$ and
$0.9$ (a few percent larger values
are obtained in the lattice simulations of Ref.~\cite{Okamoto:1999hi})%
\footnote{The lattice results for QCD that are
  frequently referred to are usually results with quarks which have
  not been continuum extrapolated and so typically give somewhat
  smaller numbers. Reliable continuum extrapolations have so far been
  done only for purely gluonic QCD. According to
  Ref.~\cite{Karsch:2000ps} the continuum extrapolation of the lattice
  results with quarks can be expected to increase the measured values
  to bring them close to the pure glue ratio.}.
Taking the Pad\'e approximant (\ref{R44}) for $\mathcal S/\mathcal
S_0$ as plausible estimate for the true nonperturbative result in
$\mathcal N=4$ SYM, a comparison of the two theories then leads
one to map the regime $T\ge 3T_c$ of pure glue QCD to couplings
$\lambda_{\rm SYM} \le 2.5$. At such values neither the 
(strict) weak-coupling nor the strong-coupling expansion is
giving reasonable approximations, however our `NLA'
resummation of perturbation theory is remarkably close to the
Pad\'e results. (Looking only at the strong-coupling result to
order $\lambda^{-3/2}$ one would be led instead to couplings
\cite{Gubser:2006qh} $\lambda \sim 5.5$ and thus closer to a
direct identification of $\lambda_{\rm QCD}$ and $\lambda_{\rm
SYM}$. However, for such a value of $\lambda$, the NLO
result in the strong-coupling expansion deviates already
significantly from the LO corresponding result of $3/4$, thus
suggesting that one should expect large corrections from higher
orders, as indeed suggested by the Pad\'e results.)

If $\mathcal N=4$ SYM theory really provides some guidance here,
this would indicate that, above $3 T_c$, the QCD coupling is not large
enough to drive the system into a regime that resembles that of the
strong-coupling limit of SYM theories.
This value of the coupling is admittedly too large to validate
naive perturbative calculations, but not so large as to prevent the
simple reorganizations of the perturbative expansion that we have
presented here, and which
seems to imply that the entropy of both theories for the corresponding
couplings can be accounted for
in terms of effectively
weakly coupled (HTL) quasiparticle degrees of freedom.

Evidently, there is important additional nonperturbative physics at
couplings $\lambda\gtrsim 4$ in SYM as there is in QCD for $T\lesssim 2.5 T_c$,
though in these respective
regimes the behavior of the entropy in the two theories is hardly 
comparable.

\acknowledgments

A.R.\ would like to thank Guy Moore for stimulating discussions.
This work was initiated during the program ``From RHIC to LHC: Achievements
and Opportunities'' at the
INT, Washington University, Seattle; E.I.\ and
A.R.\ are grateful to the INT for hospitality.
J.-P. B. thanks the Yukawa Institute for Theoretical Physics at Kyoto University, where this work has been completed
during the Yukawa International Seminar (YKIS2006)  ``New Frontiers in QCD''.
Furthermore we thank Axel Buchel and Steven Gubser for correspondence on
the correct definition of the coupling in the AdS/CFT result.


\providecommand{\href}[2]{#2}\begingroup\raggedright\endgroup

\end{document}